\documentclass[12pt,preprint]{aastex}

\shorttitle{A Ring Shaped Star Cluster}
\shortauthors{Kumar et al.}

\begin{document}


\title{A Ring Shaped Embedded Young Stellar (Proto)Cluster}


\author{M. S. Nanda Kumar\altaffilmark{1}, D. K. Ojha\altaffilmark{2}
 and C. J. Davis\altaffilmark{3}}


\altaffiltext{1}{Centro de Astrofisica da Universidade do Porto, Rua 
das Estrelas, s/n, 4150-762, Porto, Portugal}
\altaffiltext{2}{Tata Institute of Fundamental Research, Homi Bhabha 
Road, Mumbai, India}
\altaffiltext{3}{Joint Astronomy Center, 660 N. A'oh\={o}k\={u} Place,
University Park, Hilo, HI 96720, USA}


\begin{abstract}

We   present   sub-arcsec   (FWHM$\sim$0.5$\arcsec$)  J,   H,   K  and
L$^{\prime}$  images  of a  young  stellar  cluster associated with  a
candidate massive protostar IRAS\,22134+5834.  The observations reveal
a centrally  symmetric,  flattened cluster  enclosing a  central  dark
region.  The  central  dark region  is  possibly  a  cavity within the
flattened cluster. It is  surrounded by a   ring composed of  5 bright
stars   and  the candidate  massive   protostar IRAS\,22134+5834.   We
construct JHKL$^{\prime}$ color-color  and HK color-magnitude diagrams
to identify the young   stellar  objects and estimate their   spectral
types.  All the  bright stars in the ring  are found to have intrinsic
infrared  excess emission and  are likely to be early  to  late B type
stars.  We estimate an average foreground extinction to the cluster of
A$_v$$\sim$5~mag and individual extinctions to the bright stars in the
range  A$_v$$\sim$20-40~mag indicating    possible cocoons surrounding
each  massive star.  This ring of  bright stars is  devoid  of any HII
region. It is surrounded by an embedded cluster making this an example
of a  (proto)cluster that is in one  of  the dynamically least relaxed
states.   These   observations  are    consistent   with the    recent
non-axisymmetric  calculations of Li \&  Nakamura,  who present a star
formation scenario in which a magnetically subcritical cloud fragments
into  multiple   magnetically  supercritical  cores, leading    to the
formation of small stellar groups.

\end{abstract}





\section{Introduction}

Star formation is traditionally classified into isolated and clustered
modes and the majority of stars that  contribute to the galactic field
population are thought to form via the clustered mode.
\citet{am01}(Hereafter AM01) have recently demonstrated the importance
of distinguishing large  clusters (Ex: the  Orion Nebula Cluster) from
stellar groups and associations .  These authors define stellar groups
as an independent entity with the number of stars N$_{\star}$$\sim$100
and show that such groups have dynamical relaxation times much smaller
than their formation times.   Such  short dynamical relaxation   times
imply  that  small groups form and   disperse quickly after the parent
cloud gas removal, so the majority of these stellar groups are thought
to  be  overlooked in observational  open  cluster surveys (see AM01).
AM01 also  suggest  that stellar groups  and  small clusters with  the
number  of stars N$_{\star}$$<$100-300 may contribute  to  90\% of the
galactic  field population and state  that many or  most stars form in
systems with 10$<$N$_{\star}$$<$100.

In  the  last  few years  there  has   been considerable   progress in
theoretical  models    and   numerical  simulations of   multiple-star
formation  \citep{kless98,linak02}.  These simulations predict several
physical conditions and parameters that  are yet to be observationally
verified.   In  particular,   \citet{linak02}(Hereafter     LN02)  and
\citet{li01}  predict   the  formation  of   ring like   structures in
magnetically subcritical   clouds leading to  multiple-star formation.
Finding a good example of ring like structures in star-forming regions
is  difficult  because   of   the high   degree  of   geometrical  and
projectional symmetry that  is needed.  \citet{kumar02} discovered one
such stellar cluster with a high degree of geometrical symmetry around
IRAS\,22134+5834, a  putative  massive  protostar.  In  this paper  we
present    infrared  photometric   observations    of   this symmetric
ring-shaped  stellar cluster, which is all   the more unique given its
``central  dark patch''.  We shall use  these observations to evaluate
some general properties of the cluster and estimate the spectral types
of the young stellar population.

\section{Observations and Data Reduction}

All observations presented here were made on the nights of June 25/26,
2002 with the 3.8~m United Kingdom  Infrared Telescope(UKIRT) on Mauna
Kea, Hawaii. J, H and K band observations were made using the facility
imager UFTI which  is  equipped with a 1024$\times$1024  HgCdTe Hawaii
array. The plate scale was  0.09$\arcsec$/pixel with  a field of  view
(FOV) of    90$\arcsec$. The  average    seeing  in  the   K band   was
0.46$\arcsec$.    L$^{\prime}$   images were    obtained using UKIRT's
secondary imager  IRCAM, which has a  256$\times$256 InSb array with a
plate scale of  0.08$\arcsec$/pixel.  The chop/dither pattern utilized
for  the observations resulted  in a  FOV  of 38$\arcsec$ covering the
central  region  of the  cluster.  Standard  data reduction procedures
were followed, involving  dark  subtraction, division  by  a flat  and
subtraction of a  sky frame.  We  used the DAOPHOT  package in IRAF to
extract     J,H,K \& L$^{\prime}$ magnitudes      from the data.   The
instrumental magnitudes  were calibrated  using observations  of UKIRT
faint standards (FS29,FS35 \& FS149) \citep{haw01} which were observed
at  air masses   closest to  the target   observations.  The resulting
photometric data are in the natural system of the Mauna Kea Consortium
Filters \citep{st02}.  For  the   purposes of plotting these   data in
Fig.2 and Fig.3,    we have converted them  to   the \citet{bb88} (BB)
system   since the main sequence  references   are in  BB system.  The
completeness limits of the  images were evaluated by adding artificial
stars  of different  magnitudes  to  the  images and determining   the
fraction of stars recovered in  each magnitude bin.  The recovery rate
was greater than 90\% for magnitudes brighter than 18, 17.5, 17 and 12
in the J, H, K  and L$^{\prime}$ bands respectively.  Our observations
are complete(100\%) to the level of 16, 15.5,  14 and 11 magnitudes in
J,H, K and L$^{\prime}$ respectively.

\clearpage
\begin{figure*}
\plotone{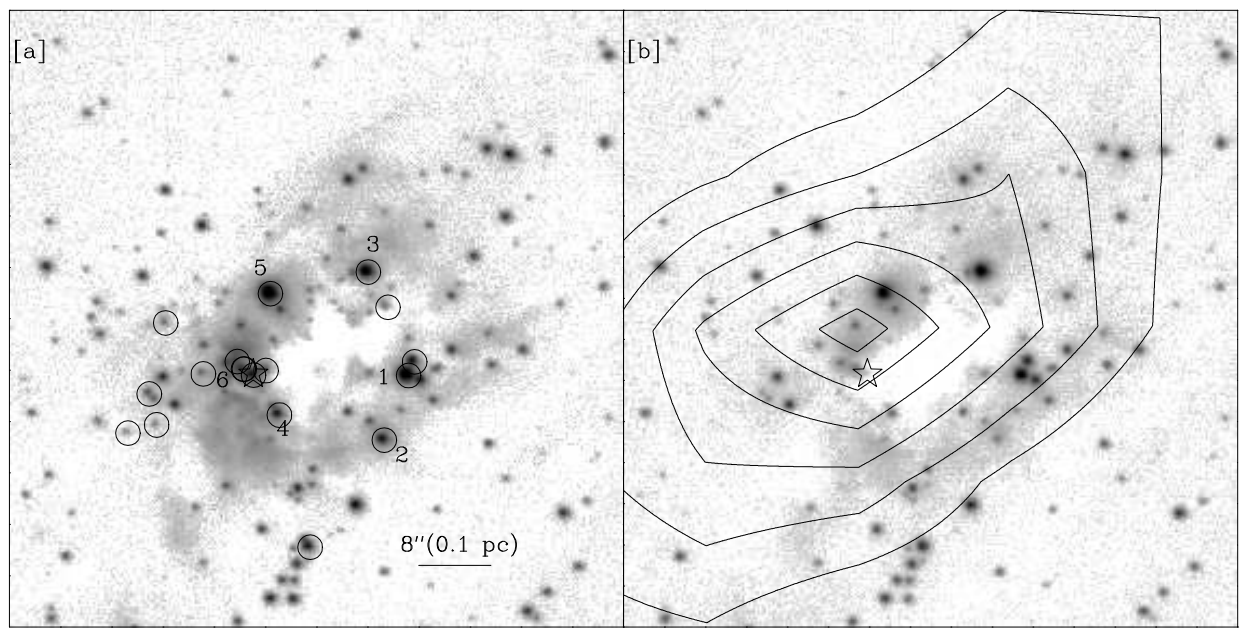}

\caption{a)A K-band image of IRAS\,22134+5834 displayed with a log scale. 
The    star   symbol    represents   the  FIR/sub(mm)   emission  peak
\citep{beu02}.   The circles  mark  all stars detected  in the  L-band
image.  The     labels designate the  identifications  of   stars from
Table.~1.  b) H-band  image of IRAS\,22134+5834 overlayed by C$^{18}$O
contours  from \citet{du01}.  Contour levels  start at 3~K~km s$^{-1}$
and increase in steps of 0.5~K~km s$^{-1}$.}

\end{figure*}

\begin{figure}
\plotone{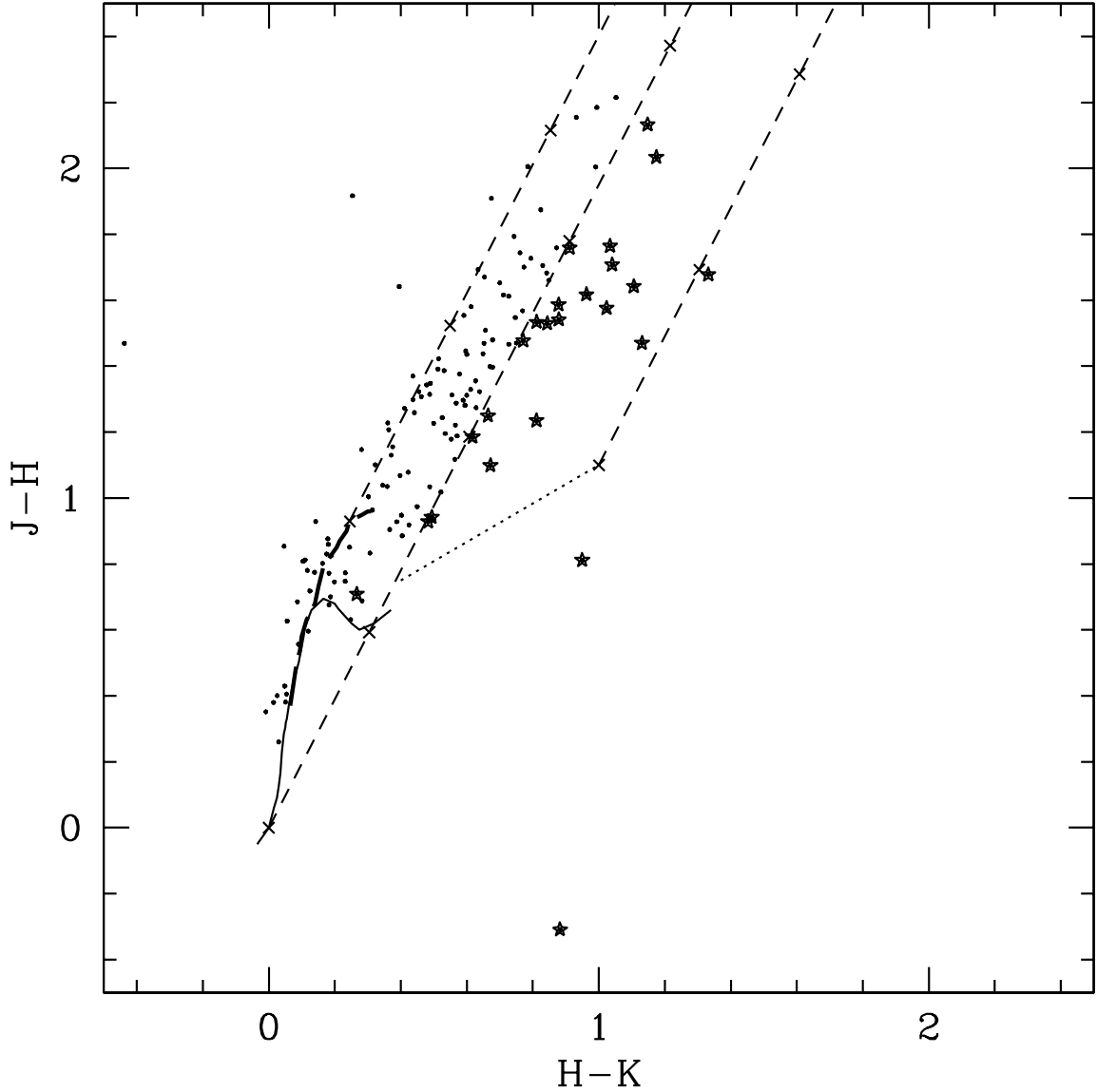}

\caption{Color-Color diagram for the 144 stars detected in the JHK bands. 
We indicate sequences for field dwarfs  (solid curve) and gaints (thick
dashed curve) from \citet{bb88}.  Dashed  straight lines represent the
reddening vectors.  The Crosses on  the dashed lines are separated by
A$_v$=5~mag.  The  dotted line represents  the  locus of T-Tauri stars
\citep{meyer97}.  Sources   that lie   to    the right(red)   of   the
main-sequence    reddening    vectors     are marked      with    star
symbols. \label{fig2}}

\end{figure}

\begin{figure}
\plotone{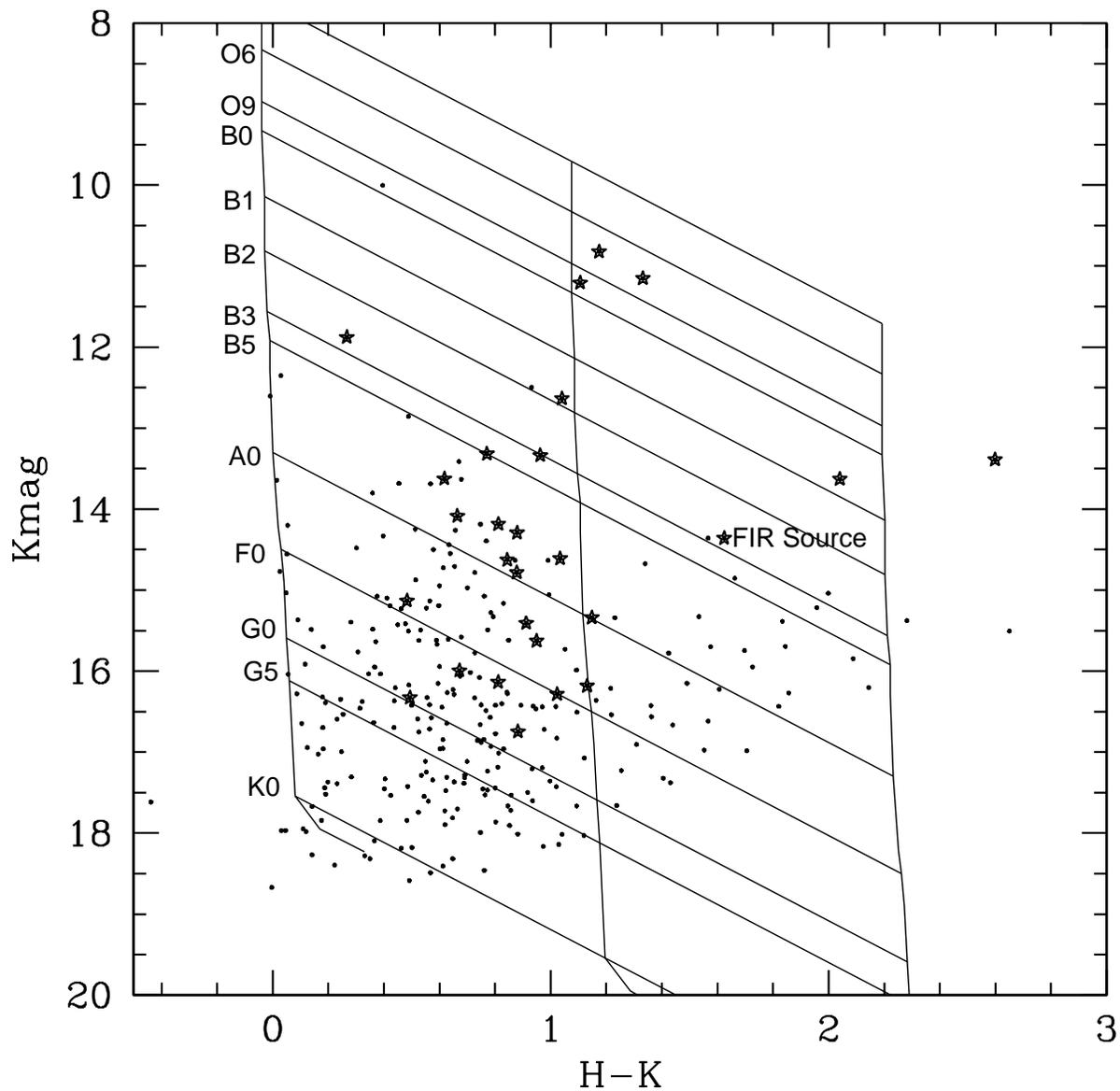}

\caption{Color-Magnitude diagram for the 270 stars detected in the 
HK  bands.  Stars  represent the  YSOs  identified  from  Fig.~2.  The
vertical solid  lines  from  left  to   right indicate   the track  of
main-sequence dwarfs reddened by 0, 20 and 40 magnitudes respectively.
The  slanting horizontal lines identify   the  reddening vectors.  FIR
source refers  to the star  associated  with the  FIR peaks  marked in
Fig.~1.\label{fig3} }
\end{figure}
\clearpage

\section{IRAS\,22134+5834: A ring like star cluster enclosing a dark patch}

IRAS22134+5834   is    a    luminous    Far-infrared   (FIR)    source
(1.7$\times$10$^3$L$_{\odot}$)  that is characterized  as  a candidate
precursor  to  an  ultra-compact HII   region \citep{sri02}.  Situated
inside  a  compact molecular  cloud    at  a  distance  of    2.6\,kpc
\citep{sri02}, this source is also  known to drive a massive molecular
outflow \citep{du01}.  FIR maps of this region at 450, 850, 1300$\mu$m
\citep{chini01}  and at    1.3mm \citep{beu02},  all reveal extended 
elliptical flux distributions with a single  resolved peak centered on
the  IRAS  source.    Figure.~1a  shows   our   K-band image of    the
target. This reveals a rich,  centrally symmetric embedded cluster and
a central dark region  associated with the  FIR/(sub)mm emission.  The
star symbol in Fig.~1 represents the position of the FIR/(sub)mm peaks
and  coincides  with  a    star  visible  at wavelengths longward   of
1.6$\mu$m.    Although the FIR  emission  encircles  the infrared dark
region, there are   no signatures  of  any  condensations within   the
extended emission.   Note the faint stars  that delineate the northern
triangular shaped boundary of the dark  region.  The neat alignment of
the stars  along this boundary  suggests that  the central dark region
cannot be  caused  by a foreground   object.   Assuming that the  dark
region is an integral part of the  cluster implies that the cluster is
flattened in   a  plane   perpendicular to  the   line   of  sight.  A
spherically  symmetric   cluster would have  produced    an image with
several stars seen  in projection against  the dark patch.  Thus,  the
image in Fig.~1 probably represents the true distribution of stars and
cannot result  from  an arbitrary  3D distribution.  The  bright stars
surrounding  the dark patch in  the image must  be arranged  in a true
ring-like pattern around   the dark patch.   However, for  comparision
with cluster formation models we  must also establish whether the dark
region is an empty cavity or a molecular core?

It can be seen from  the overlay of  C$^{18}$O contours on the  H-band
image (see Fig.~1b) that the dark region  is immersed in the C$^{18}$O
emission;   both the  C$^{18}$O   emission and   the dark region   are
elongated approximately  in  an  east-west direction  with  a  similar
aspect ratio.  \citet{du01}  estimated a total mass of  206M$_{\odot}$
for this core and a   velocity  gradient of 0.9~km   s$^{-1}$pc$^{-1}$
along the  core's   major axis.  \citet{rich87}   detected strong HCO+
emission  from  this   source,  confirming  the presence   of  a dense
molecular cloud. \citet{sri02} detected  NH$_3$ (J,K)= (1,1) and (2,2)
lines and  measured a rotation  temperature  of 18K  for this  region.
However, the peaks of these molecular  line emission regions, like the
FIR  emission peaks, are  all   centered on IRAS\,22134+5834 which  is
situated on the periphery of  the ring of  stars. These dust continuum
and dense molecular tracers are {\em not} centered  on the dark patch.
These results favour the idea that the dark region  is an empty cavity
rather than a dense core.  There is, however,  one fact which suggests
that the dark region is a dense core; the faint stars in Fig.~1a which
delineate the northern  triangular shaped boundary  of the dark region
are visible only in the K-band  and not in  the H (Compare Fig.~1a and
1b) or J bands.  These stars could  therefore be deeply embedded along
the boundary of  this dark region.   Nevertheless, conclusive proof as
to whether the dark patch is a cavity or  a molecular core can only be
obtained from a higher-resolution  map of the  dense molecular gas  in
this region.

\subsection{Photometric Analysis}

Inside   a   100$\arcsec$$\times$100$\arcsec$  field    centered    on
IRAS\,22134+5834 we found 145 stars in J, 286 stars in H and 357 stars
in K,  with magnitude errors less than  0.2.  Of these, 144  stars are
found to be common to all three JHK bands  and 270 stars are common to
the HK bands  only.  Note that almost all  the stars detected in the J
and H bands are also detected in the  K-band.  Since many of the stars
are detected in the K  band and not in  the J band, the cluster itself
must be deeply embedded in the parent molecular  cloud.  Thus a H-K vs
J-H color-color (CC) diagram can  not completely identify the embedded
young stellar population.  However, in Fig.~2 we show a CC diagram for
the 144  stars detected in the JHK  bands.  The solid and broken heavy
curves represents   the    main-sequence  dwarf and    giant    stars,
respectively, and the dashed  parallel lines are the reddening vectors
that  enclose  reddened   main-sequence  objects.   The   dotted  line
indicates  the locus of T-Tauri  stars \citep{meyer97}.   25 stars lie
outside the region of reddened main-sequence  objects; these are young
stellar objects (YSOs) with intrinsic color excesses .  By dereddening
the stars (on the CC diagrams)  that fell within the reddening vectors
encompassing  the main sequence stars and  giants, we found the visual
extinction to each star. We deredenned the stars  to the K6-M6 part of
the sequence of stars.  The individual extinction  values range from 0
to 14 magnitudes.  From a  histogram of these  values, we estimate the
average foreground extinction  to be A$_v$$\sim$5~mag.  Similarly, for
sources detected in the H, K and L$^{\prime}$ bands, we constructed an
K-L$^{\prime}$ vs H-K CC diagram.  Among the 11  stars detected in the
H,  K and L$^{\prime}$ bands,   7 stars were  found  to  be YSOs  with
intrinsic   color   excesses.   Four of    them show   extinctions  of
A$_v$$\sim$20-40 mag,  indicating that  the individual extinctions  to
the cocoons that contain these stars are much  higher than the average
extinction through the molecular cloud  that hosts the cluster.  These
four  stars are among  the bright stars that form  the ring around the
central dark patch. We note that all the  bright stars surrounding the
central dark  patch in Fig.~1 are  also seen in the L$^{\prime}$ image
and are found to have an intrinsic infrared excess.
 
Figure.~3 shows a H-K  vs K color-magnitude  (CM) diagram for all  the
sources detected in the H \& K bands.   YSOs found from the CC diagram
(Fig.~2) are shown as star symbols.  However, it  is important to note
that even  those stars not  shown with a  star symbol may also be YSOs
with an intrinsic  color excess.  The  vertical solid lines (from left
to right)  represent the main-sequence curve  reddened by 0, 20 and 40
magnitudes respectively.  We have assumed a distance of 2.6~kpc and an
A$_v$$\sim$5~mag to the source to  reproduce the main sequence data on
this plot.  The CM diagram is a useful  tool for estimating the nature
of the stellar  population within the  cluster  in the absence of  any
spectroscopic data.   However, it  can  be  highly misleading  if  the
sources have intrinsic color excesses. In the pretext of this work, it
is of interest to obtain a census of massive stars in the region.  The
horizontal slanting lines in Fig.~3 trace the reddening zones for each
spectral type.  Since the sources we  are interested in have intrinsic
infrared  excesses, the spectral   type estimation will be misleading.
We  therefore considered the  candidate massive protostar (FIR source)
IRAS\,22134+5834, as the reference point  on the CM diagram (marked in
Fig.~2) to  identify other massive stars.   It can be  seen that there
are    6-8 points  lying either   above    or in  the    same range as
IRAS\,22134+5834.   These stars are   identified on  the  K-band image
(Fig.~1a) and are found to constitute the ring of  bright stars.  This
suggests that  there are at least four  stars  (appearing more massive
than IRAS\,22134+5834)  and two   stars (similar to  IRAS\,22134+5834)
that are all situated in a ring enclosing the central  dark patch.  We
also estimated the spectral types of the same sources on a J-H vs H CM
diagram and  obtained similar results  thus verifying the consistency.
Table.~1 lists the positions, individual  extinctions, fluxes in the K
and L$^{\prime}$  bands  and   corresponding flux  ratios   indicating
approximate colors.   The candidate massive protostar IRAS\,22134+5834
is also situated  along the periphery of the  ring and is estimated to
be of spectral type B3.  The luminosity of this FIR point source (logL
= 3.23)  indicates a  luminosity equal to   that  of a  Zero Age  Main
Sequence (ZAMS) star of B2-B3 \citep{pan73}.

\section{Discussion}

The results of sections 2  and 3 show that IRAS\,22134+5834 represents
an  embedded young star cluster  associated with a luminous FIR source
but {\em not} with an HII region.  The  K-band image reveals 208 stars
in the central 1~pc region (slightly larger than the area displayed in
Fig.~1).   If   we consider incompleteness  in    sampling, and volume
filling factors, this cluster  represents an upper  limit example of a
cluster with a number of stars N$_{\star}$$<$100-300, which, according
to AM01 is a small cluster, a type that contributes to the majority of
the   galactic field population.     The  embedded  nature,  flattened
appearance of the cluster and the  ring of bright stars, together with
the lack  of a significant  HII region, demonstrates the extreme youth
of this small    cluster.  It thus    makes an  excellent  target  for
verifying  predictions of  the  theories  of cluster formation.    The
standard  scenario   for  the  formation of   isolated  low-mass stars
involves an axisymmetric cloud   evolution leading to a  single  dense
core.   Recently LN02 presented  calculations of  the non-axisymmetric
evolution of   a  magnetically   subcritical molecular   cloud   under
thin-disk approximation  (flattened cloud) fragmenting  into  multiple
magnetically super-critical cores.  Such  calculations are believed to
be  fundamental to the  formation of  all varieties  of star formation
including singles, binaries,  multiples and  clusters.  These  authors
predict that the supercritical  cores resulting from fragmentation are
arranged in a ring shape because  the magnetic field tension prohibits
the formation of a  central  singularity.  The observations  presented
here tend to support their predictions since  the flattened cluster of
IRAS\,22134+5834 shows a ring of young massive stars which has not yet
formed an HII region.  Further, the central dark patch appears to be a
cavity rather than a core (\S.~3), as  predicted by LN02.  Indeed, the
K-band image  of Fig.~1  is  strikingly similar to  the  numerical
simulations of LN02 for an arbitrary case  of a perturbation with mode
m=5  (Fig.~3c of LN02).  Rings  of stars at  the center of a molecular
cloud containing  several   jeans  masses can  also   be  seen in  the
numerical   simulations  of   \citet{kless98}.  The  estimated average
foreground extinction  to   the cluster  is A$_v$$\sim$5~mag and   the
individual   extinctions      to   the stars       are    as  high  as
A$_v$$\sim$20-40~mag.  This indicates that the massive stars composing
the ring in   Fig.~1 are surrounded by  independent   cocoons that may
represent individual dense cores.

While the  CM diagram estimates  demonstrate  that there are at  least
five stars in the   cluster with masses similar to   IRAS\,22134+5834,
there is no  significant HII region  associated with  this source. The
intrinsic colors of  massive stars are  not well  understood and a  CM
diagram  takes into account  only reddening  and not intrinsic colors.
Thus any   star with intrinsic  color  would move  not  only along the
reddening line on  a CM diagram but also   upwards (along the  y-axis)
resulting in an overestimation of the mass.  To evaluate the magnitude
of such an overestimation we  placed the candidate low mass protostars
in  Taurus   on CM diagram    by  taking the  K   and  H-K  data  from
\citet{pk02}.  These Class I or  Class 0 low mass protostellar sources
were found  to be distributed  between spectral  types G5 and  B5.  We
find that a  1-2~M$_{\odot}$ star can be  mistaken for a 6~M$_{\odot}$
star.  Exactly how this behaviour transforms  at higher masses can not
be   judged  with this  simple exercise.    We would   need a detailed
understanding  of the intrinsic colors  of the massive young stars and
some identification of their exact protostellar phases. In the present
context we can only state that there are at least 5 stars with similar
or higher masses than the candidate protostar IRAS\,22134+5834.  While
the candidate protostar is  bright at FIR  wavelengths and not  at NIR
wavelengths, the remaining 4 stars are  bright at 2$\mu$m.  Weak radio
free-free emission is detected at 3.6~cm centered within 1$\arcsec$ of
the FIR source ( Kurtz,  S.  Pvt Communication);  this is also evident
in the NRAO VLA Sky  Survey data.  In the  light of the above results,
it is surprising that even this weak radio emission is centered on the
FIR source and {\em not} on the other bright stars which are estimated
to have masses greater than the FIR source.  Further investigations to
establish   the clear  association of   these  bright  stars with  the
molecular gas, and spectroscopic studies  to infer the exact  spectral
types of the stars, can resolve the issue.   We summarize this work by
noting that the cluster is associated with a ring of candidate massive
young stars.  While rings of  massive stars are  not a new  phenomenon
(ex: W49A \citet{welch87}) ,  a ring of  candidate massive young stars
devoid of ionized regions is relatively unique.  IRAS\,22134+5834 thus
represents  an  early  phase  in  massive   star  formation   when the
associated  cluster is in one of  the dynamically least relaxed states
known.

\section{Summary}

Infrared  photometric  studies of  an  embedded young  stellar cluster
associated  with IRAS\,22134+5834 are  presented.
\noindent 1) The cluster is centrally symmetric, flattened 
and encloses a central dark region that appears to be an empty cavity.
This dark region is surrounded by a ring of bright stars.
\noindent 2) The ring of bright  stars are estimated to be composed
of  four stars   likely  to  be more  massive   than the  FIR   source
IRAS\,22134+5834   and   two stars   with similar  masses   to that of
IRAS\,22134+5834.  However, the  cluster does not have any  associated
HII region, implying its  extreme youth.  While an average  foreground
extinction to   the cluster  is   estimated  to  be  A$_v$$\sim$5~mag,
extinction to the  individual bright stars in the  ring are as much as
A$_v$$\sim$40~mag indicating  cocoons of  dense  gas surrounding  each
star.  
\noindent 3) The results are consistent with the calculations of 
LN02   who  present a  star     formation scenario in  a  magnetically
subcritical cloud fragmenting into multiple magnetically supercritical
cores  leading to the formation of  small stellar groups.  The central
dark patch in  IRAS\,22134+5834 and the   surrounding ring of  massive
stars display   striking   similarity  to the  numerical   simulations
presented by these authors.

\acknowledgments

We   thank Mario  Tafalla  for useful  comments   and K.   Dobashi for
providing the original C$^{18}$O data.

\clearpage

\begin{deluxetable}{ccccccc}
\tabletypesize{\scriptsize}
\tablecaption{Bright Stars constituting the Ring}
\tablewidth{0pt}

\tablehead{ \colhead{ID} & \colhead{R. A J2000}& \colhead{DEC J2000} & \colhead{A$_v$} & \colhead{F$_{\lambda}$(K)} & \colhead{F$_{\lambda}$(L$^{\prime}$)} & \colhead{$\frac{\lambda F_{\lambda}(L^{\prime})}{\lambda F_{\lambda}(K) }$} \\
\colhead{(from Fig.1)} & \colhead{hh:mm:ss} & \colhead{dd:mm:ss} & \colhead{mags} & \colhead{Jy} & \colhead{Jy} & \colhead{} \\}

\startdata

1 & 22:15:6.8 & 58:49:07 &  12.1 & 0.0794  & 0.0577  & 1.2476 \\
2 & 22:15:7.1 & 58:49:00 &  38.4 & 0.1532  & 0.0490  & 0.5492 \\
3 & 22:15:7.3 & 58:49:18 &  $\sim$0 & 0.0116  & 0.0032  & 0.4674 \\ 
4 & 22:15:8.6 & 58:49:02 &  26.7 & 0.0363  & 0.0167  & 0.7910 \\ 
5 & 22:15:8.7 & 58:49:16 &  13.9 & 0.1292  & 0.0655  & 0.8695 \\ 
6 & 22:15:9.1 & 58:49:07 &  5.4  & 0.0021  & 0.0041  & 3.3918 \\

\enddata

\end{deluxetable}


\end{document}